\documentclass[runningheads]{llncs}
\usepackage[T1]{fontenc}

\usepackage{graphicx}

\usepackage{multirow}
\usepackage{amsmath}

\begin{document}
\title{Semi-Supervised Speech Confidence Detection using Pseudo-Labelling and Whisper Embeddings}
\titlerunning{Semi-Supervised Speech Confidence Detection using Whisper Embeddings}

\author{Adam Wynn\inst{1}\orcidID{0000-0002-1631-2151} \and
Jingyun Wang \inst{1}\orcidID{0000-0001-9325-1789} \and
Xiangyu Tan \inst{2}\orcidID{0000-0001-9171-1788}}
\authorrunning{A. Wynn et al.}
\institute{Durham University, Durham, United Kingdom \email{\{adam.t.wynn, jingyun.wang\}@durham.ac.uk}
\and Shanghai Open University, Shanghai, China \email{tanxy@shisu.edu.cn}}

\maketitle              

\begin{abstract}
\vspace{-3mm}
Understanding speaker confidence is crucial in educational settings, as it can enhance personalised feedback and improve learning outcomes. This study introduces a novel framework for detecting speaker confidence by integrating human-engineered features with embeddings from the Whisper encoder. To address data limitations, a pseudo-labelling technique is employed to expand the labelled dataset, allowing the model to learn from both human-annotated and model-generated labels. The framework combines traditional speech features including pitch, volume, rate of speech, and the presence of disfluencies and stress, with  Whisper embeddings, and uses a co-attention mechanism to fuse these representations and achieve an overall accuracy of 75\%. This study contributes to advancing speech analysis, enabling applications that support personalised learning and speaking skill development.

\keywords{Confidence Detection \and Disfluency Detection \and Semi-Supervised Learning \and Speaking Skills}
\end{abstract}

\section{Introduction}

Effective communication is essential in education, shaping how knowledge is shared and understood \cite{Kasemsap2017}. Confidence, in particular, influences a student’s clarity, credibility, and engagement when speaking \cite{Mardiana2024}, making it vital for tasks like presentations and public speaking. Understanding confidence can support automatic feedback and help  educators identify areas where support is needed \cite{Cavalcanti2021} to help improve communication skills.

Prior research has demonstrated that confidence is reflected in acoustic features such as pitch, speech rate, and vocal intensity \cite{Jiang2014}. However, earlier work relied on manual annotation and small-scale studies \cite{Williams2017}, and despite advances in AI \cite{Nair2020}, confidence detection remains underexplored due to a lack of annotated datasets. Therefore, this study explores the detection of speaker confidence, by proposing a novel semi-supervised framework that integrates human-engineered features including pitch and speech rate, with embeddings from the Whisper encoder \cite{Whisper} using a co-attention mechanism. Human-engineered features are used to generate pseudo-labels \cite{psuedolabel} for unlabelled data to expand the dataset and improve the generalisability of the model.

This work aims to answer the following research questions: \textbf{RQ1:} How can semi-supervised learning and model-based pseudo-labelling be leveraged to address the scarcity of confidence-labelled data?  \textbf{RQ2:} To what extent do pitch, rate of speech, amplitude and the presence of disfluencies and stress, influence the model's perception of confidence? Our main contribution is introducing a semi-supervised framework to detect speaker confidence, offering a solution for assessing students’ verbal communication skills and enabling adaptive, real-time feedback in educational settings.

\section{Related Work}
\vspace{-5mm}

Confidence in speech is generally perceived through a combination of vocal characteristics including pitch, volume, speech rate, and clarity. These features act as the foundation for evaluating speaker confidence in both human and automated systems.  Prior research \cite{JIANG2017106} discovers that confident expressions have highest f0 range, mean amplitude and amplitude range and unconfident expressions are highest in mean f0, slowest in speaking rate, with more frequent pauses.
Automated systems designed for confidence detection have been proposed including providing automated feedback to users rehearsing oral presentations based on speech quality, content coverage and audience reaction \cite{Trinh2017}, and predicting the confidence of a speaker using Mel-Frequency Cepstral Coefficients (MFCC) as inputs based on clarity, modulation, pace, and volume. 

Confidence can also influence the frequency of speech disfluencies and a lack of confidence often results in more frequent pauses and fillers \cite{astuti2024speech}. Several approaches for disfluency classification have recently emerged \cite{Kourkounakis20}\cite{Boughariou2024}, but require large labelled datasets. To address this challenge, Mohapatra et al. \cite{Mohapatra2022} proposed DisfluencyNet, a Wav2Vec 2.0-based model with convolutional and fully connected layers, achieving over 5\% improvement in disfluency detection compared to baselines with only a few minutes of data. Moreover, Ameer et al. \cite{WhisperDisfluency} used the Whisper model for multi-class disfluency classification, introducing an encoder-freezing strategy, outperforming Wav2Vec 2.0 models.

Confidence detection is also closely linked to Speech Emotion Recognition (SER), as both fields rely on analysing the speakers' prosody.
Pepino et al. \cite{Pepino2021} proposed using Wav2vec 2.0 embeddings for SER by combining the output of several layers of pre-trained Wav2Vec 2.0 using trainable weights to produce richer speech representations. Additionally, they integrated prosodic features into the model, including pitch and loudness, resulting in further improvements in performance. 
Goel et al \cite{Goel2024} proposed to improve the generalisability of SER models by introducing CAMuLENET and incorporating attention which outperforms baseline Whisper, Wav2Vec 2.0 and HuBERT and improves the generalisation to unseen speakers. These advancements highlight the potential of deep learning in enhancing SER and confidence detection by effectively extracting prosodic cues embedded in speech to provide more accurate and generalisable predictions.  

\section{Methodology}

\begin{figure}
\centering
\includegraphics[width=8cm]{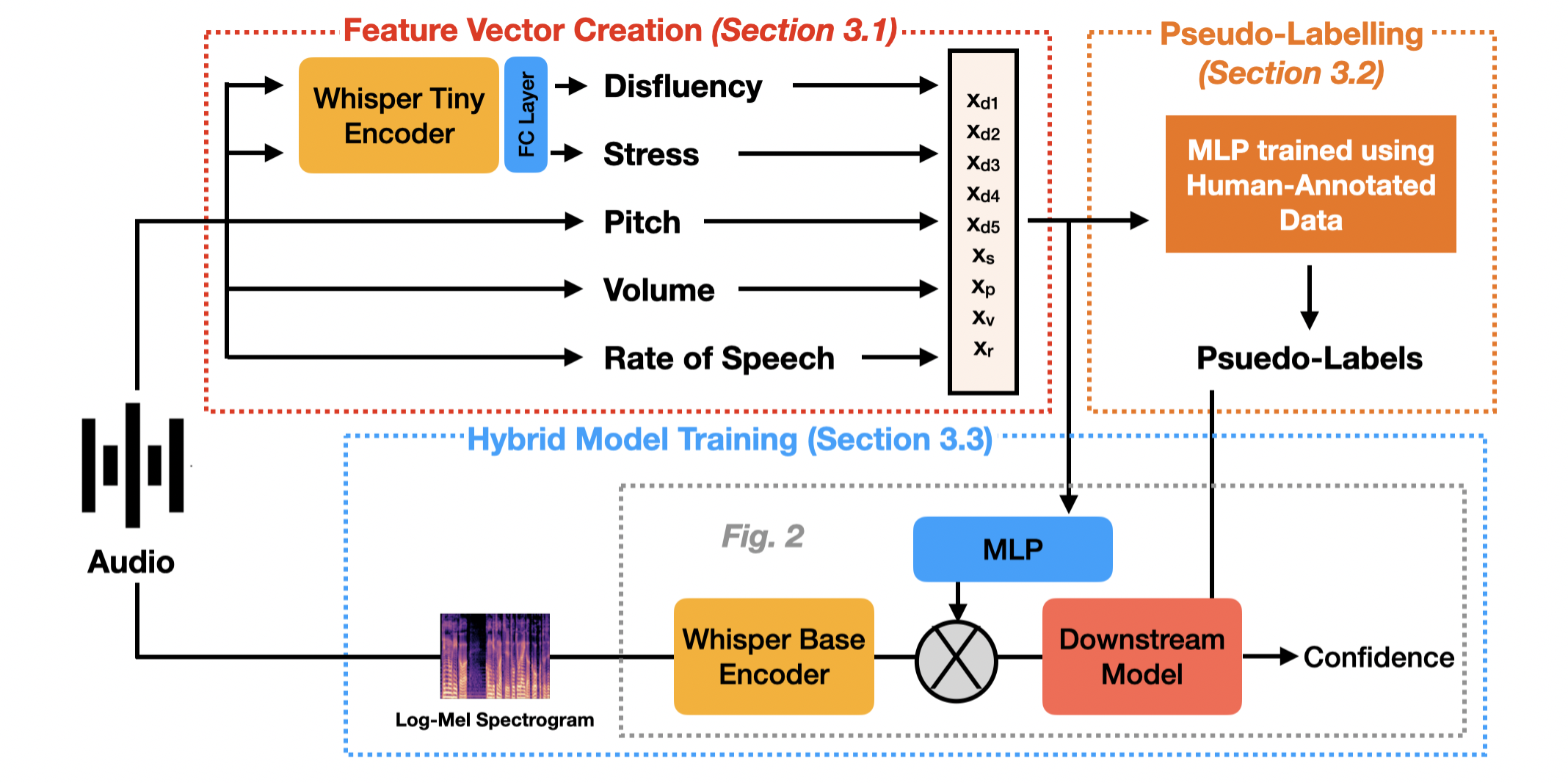}
\caption{Overall Pipeline of the Confidence Classification System during Training} \label{fig:pipeline}

\end{figure}
Building on the advancements of using deep learning for feature extraction and the classification of audio, this study proposes a novel framework for classifying speaker confidence which involves a pipeline (Fig. \ref{fig:pipeline}) that integrates feature vectors with audio embeddings using a pre-trained Whisper-base model encoder. Both representations are fused through a co-attention mechanism to incorporate information from both sources effectively and develop a robust system for classifying speaker confidence levels on limited annotated data.

Due to the lack of publicly available data labelled for confidence, a manual annotation system was developed to classify audio clips into low, medium, and high confidence. Three fluent English speakers (a native male, a non-native female, and a non-native female English speech expert) rated each clip, with final labels based on averaged scores. The resulting dataset includes 444 clips (247 male, 197 female), drawn from TEDLIUM \cite{Tedlium}, SEP-28K \cite{SEP28K}, and non-native English assessments, comprising 172 high-, 151 medium-, and 121 low-confidence samples. This dataset serves as the human-annotated test set for evaluating performance on unseen data. To compensate for its limited size, a model-based pseudo-labelling technique is used to expand the dataset for training and validation. Further details are provided in Section 3.2.

\vspace{-5mm}
\subsection{Feature Vector Creation}

A 9-dimensional feature vector was engineered for each audio clip to capture prosodic cues linked to speaker confidence, including pitch and amplitude variation, rate of speech, presence of stress, and five types of speech disfluencies: word repetitions, prolongations, interjections, blocks, and sound repetitions. Prior studies link pitch, amplitude, and speaking rate to perceived confidence \cite{JIANG2017106}\cite{Guyer2019}, whilst disfluencies and stress have been associated with low confidence \cite{astuti2024speech}\cite{stressconfidence}.

In this study, pitch variation is measured using the SPICE pitch tracker \cite{SPICE}, and amplitude is calculated using the normalised variation of the amplitude envelope, which reflects the intensity and energy of speech delivery. The rate of speech is derived using the MyProsody \cite{myprosody} library.

The Disfluency and stress detection models share a common model architecture. 
In order to identify disfluencies, data from the SEP-28K \cite{SEP28K} and FluencyBank \cite{FluencyBank} datasets is used. Only the audio clips where all three raters agreed were used and data was excluded which had poor audio quality. The dataset was augmented and balanced by applying pitch shifting and Gaussian noise to the disfluent audio files. The model architecture consists of the Whisper Tiny Encoder \cite{Whisper}, chosen due to its ability to generalise compared to the base model, which appeared to overfit the data.

 For the stress detection task, a possible solution to the lack of sufficient labelled data is to establish a mapping between emotional states and stress levels. Research indicates that the emotions sadness, fear, and anger are most closely associated with high stress levels \cite{Stavs2023}. To address this, audio segments were extracted from the RAVDESS \cite{RAVDESS}, SAVEE \cite{SAVEE} and TESS \cite{TESS} SER datasets, where the labels for sadness, fear, and anger were relabelled as "stress", while the remaining labels were labelled as "neutral".

\subsection{Model-Based Pseudo-Labelling }

To address the limitation of a small ground truth dataset, a model-based pseudo-labelling approach was used to generate additional labelled data. After the feature vectors were created, a multi-layer perceptron (MLP) was trained on the ground truth dataset feature vectors to classify speech confidence into three levels: low, medium, and high. This process did not involve the audio or the Whisper embeddings directly in an attempt to prevent data leakage. 

This model was initially trained on 363 samples with Adam optimiser \cite{AdamOpt} and cross-entropy loss, and achieved an accuracy of 79.19\% using 10 fold validation. The trained MLP was then applied to a larger, unlabelled dataset comprising 2640 audio samples (880 clips per label after downsampling to ensure equal class sizes) to generate pseudo-labels. The resulting pseudo-labelled dataset, combined with the original ground truth samples, provided a significantly larger dataset for subsequent stages of the model development pipeline.

\subsection{Hybrid Model Training}

\begin{figure}
\centering
\includegraphics[width=8cm]{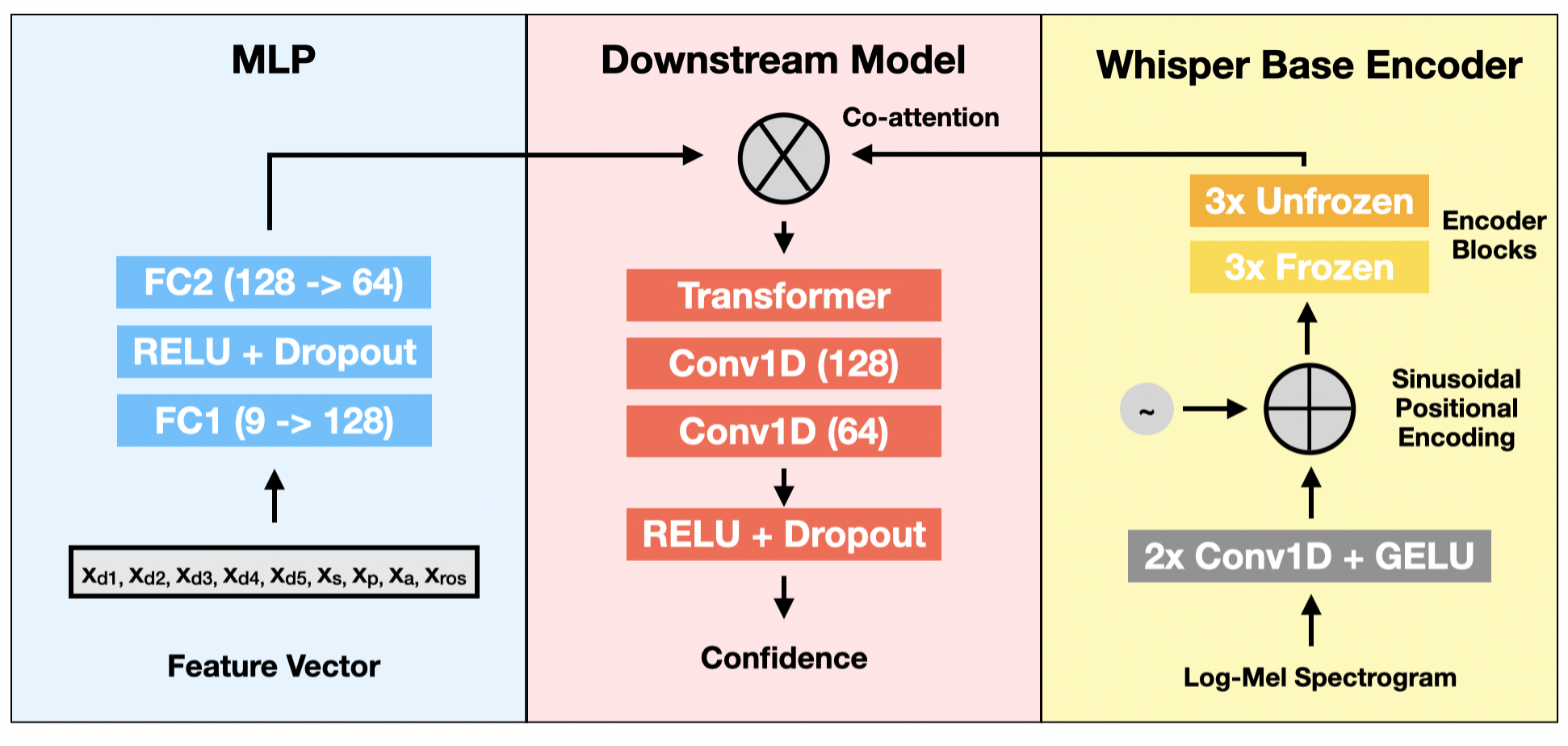}
\caption{Confidence Detection using Whisper-Base Encoder and Feature Vector} \label{fig:hybrid}
\vspace{-5mm}
\end{figure}

The hybrid model (Fig. \ref{fig:hybrid}) integrates information from both the engineered feature vector and audio embeddings derived from Whisper-base. The training process uses the pseudo-labelled dataset for training and validation, whilst the original ground truth dataset was reserved for testing.

First, the engineered feature vector was input into a MLP different from the one used during pseudo-labelling, to generate a 128-dimensional embedding. Simultaneously, the pre-processed audio data was input into the Whisper-base encoder to extract high-dimensional embeddings from the last encoder layer. These two representations were fused using a co-attention mechanism, where attention weights were applied to the Whisper embedding. The combined embeddings were passed through a downstream network to predict confidence levels.

\section{Results and Discussion}

\subsection{Disfluency and Stress Detection}

The disfluency classification model (Fig. \ref{fig:pipeline}) was trained for 50 epochs using AdamW Optimiser \cite{AdamWOpt} with a learning rate of $2.5 \times 10^{-5}$ and binary cross-entropy loss, using 10-fold cross-validation and early stopping. The model performed best on interjections (Acc: 0.80, F1: 0.80) and prolongations (Acc: 0.78, F1: 0.77), followed by sound repetitions (Acc: 0.74, F1: 0.74), word repetitions (Acc: 0.68, F1: 0.68) and blocks (Acc: 0.69, F1: 0.68). These findings align with existing literature, indicating that detecting blocks and word repetitions poses challenges \cite{Mohapatra2022} whilst interjections are more easily identified \cite{Wav2VecDisfluencyTransformer}. On the other hand, the stress classification model was trained with the same hyperparameters as above and achieved an accuracy of 0.86 (F1: 0.85).

\vspace{-3mm}
\subsection{Confidence Detection}

The Confidence classification model, as shown in Fig. \ref{fig:hybrid},  was trained for 200 epochs using the AdamW optimiser \cite{AdamWOpt}, with an initial learning rate of $2.5*10e^{-5}$ using cross entropy loss. Training and validation used data from the pseudo-labelled data and the model was tested on the human-annotated data only.

\begin{table}
\vspace{-5mm}
    \centering
    \begin{tabular}{c|cccc}
         &  Accuracy&  F1&  Precision& Recall\\
         \hline
         Low Confidence &  0.88&  0.80&  0.73& 0.88\\
         Medium Confidence&  0.61 & 0.67  & 0.74 & 0.62\\
         High Confidence & 0.78 &0.79  & 0.79 & 0.78\\
        Overall& 0.75 & 0.75 & 0.75 & 0.75\\
         
    \end{tabular}
    
    \caption{Results for the Confidence Classification Model}
    \label{tab:confidence_results}
   \vspace{-5mm}
\end{table}

\begin{figure}
\vspace{-5mm}
\includegraphics[width=\textwidth]{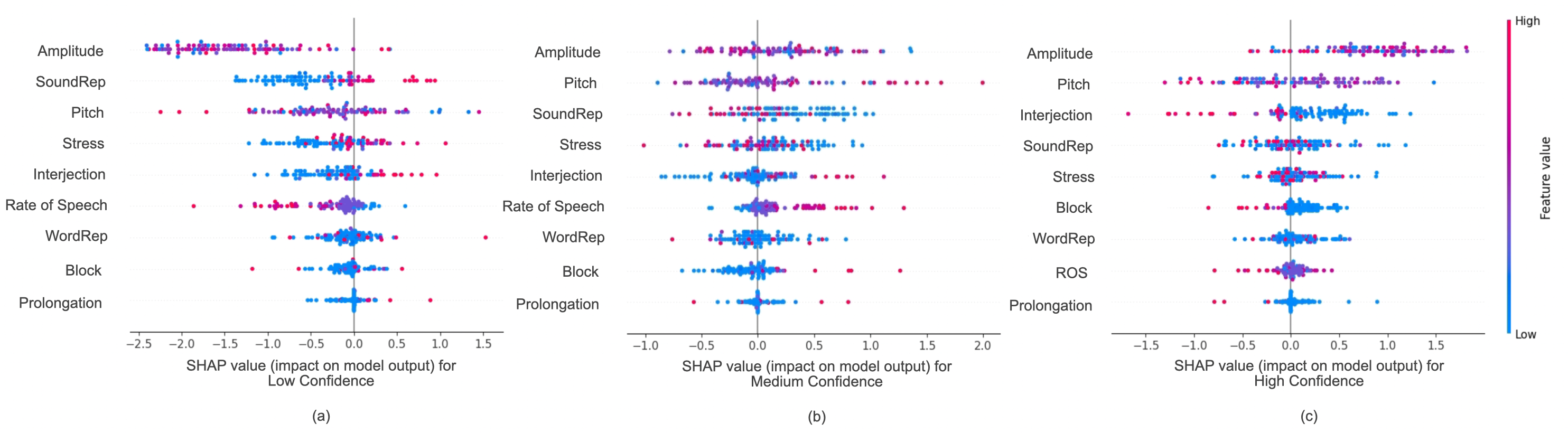}
\vspace{-5mm}
\caption{Shap Values for Low, Medium and High Confidence Respectively} \label{fig:shap}
\vspace{-5mm}
\end{figure}

As shown in Table \ref{tab:confidence_results}, the model achieves stronger results for low (Acc: 0.88, F1: 0.80) and high (Acc: 0.78, F1: 0.79) confidence, but struggles with medium confidence (Acc: 0.61, F1: 0.67), perhaps because medium confidence is more ambiguous and can be misclassified as either low or high confidence. After training the model, SHapley Additive exPlanations (SHAP) identified Pitch Variation, Amplitude Variation, and Sound Repetitions as the most important features. As shown in Fig. \ref{fig:shap}, sound repetitions strongly increased the likelihood of classifying the speaker as low confidence, though the effect on high confidence was less clear. For pitch, higher variation typically correlated with negative SHAP values for both low and high confidence, suggesting that the model associates high pitch variation with medium confidence. Amplitude variation, the most influential feature, was less clear, but in general, samples with lower amplitude variation were less likely to indicate low confidence. 

\section{Conclusion}

This study presents a novel framework for detecting speaker confidence in audio, combining feature engineering and pseudo-labelling to create a robust classification system. By training a model to generate pseudo-labels, the final hybrid model can incorporate these pseudo-labels during training to differentiate between confidence levels enabling the model to generalise its learning, resulting in stronger performance when applied to human-annotated data.

A key limitation is that the pseudo-labelling process depends on the quality of human-annotated data. Biases in this data can introduce label noise and although the hybrid model doesn’t directly use the annotations, completely unseen data would support more robust training. Enhancing the dataset with more annotators and diverse sources would also improve generalisability. Another limitation is the lack of consideration for linguistic and cultural differences in confidence expression. Further testing across languages and cultures will be necessary to ensure the model can generalise effectively across cultures.

In the future, this framework could support educational applications, such as assessing student confidence in oral presentations and providing targeted feedback. Moreover, the framework could be used to predict other abstract communication skills beyond confidence, such as persuasiveness or empathy. This study lays the groundwork for advancing speech analysis systems and paves the way for applications that allow for personalised speaking skills development.

\begin{credits}

\subsubsection{\discintname}
The authors have no competing interests to declare that are
relevant to the content of this article.
\end{credits}

\bibliographystyle{splncs04}

\end{document}